\documentclass[proof]{pasj01}
 \SetRunningHead{Astronomical Society of Japan}
 {}

\usepackage{ulem}
\usepackage{color}

\begin{document}

\title{Discovery of 6.4 keV line and soft X-ray emissions from G323.7$-$1.0 with Suzaku }

\author{Shigetaka \textsc{Saji},\altaffilmark{1, 2}\altaffilmark{*}
Hironori \textsc{Matsumoto},\altaffilmark{2}
Masayoshi \textsc{Nobukawa},\altaffilmark{3}
Kumiko K. \textsc{Nobukawa},\altaffilmark{4}
Hideki \textsc{Uchiyama},\altaffilmark{5}
Shigeo \textsc{Yamauchi},\altaffilmark{4}\\
Katsuji \textsc{Koyama},\altaffilmark{6}
}

\altaffiltext{1}{Division of Particle and Astrophysical Science,
Graduate School of Science,\\ Nagoya University,
Furo-cho, Chikusa-ku, Nagoya, 464-8602, Japan}
\altaffiltext{2}{Department of Earth and Space Science, Graduate School of Science, Osaka University,\\
 1-1 Machikaneyama-cho, Toyonaka, Osaka 560-0043, Japan}
\altaffiltext{3}{Faculty of Education, Nara University of Education, Takabatake-cho, Nara, Nara 630-8528, Japan}
\altaffiltext{4}{Department of Physics, Faculty of Science, Nara Women's University, Kitauoyanishi-machi, Nara, Nara 630-8506, Japan}
\altaffiltext{5}{Faculty of Education, Shizuoka University, 836 Ohya, Suruga-ku, Shizuoka, Shizuoka 422-8529, Japan}
\altaffiltext{6}{Department of Physics, Graduate School of Science, Kyoto University, Kitashirakawa-oiwake-cho, Sakyo-ku, Kyoto, Kyoto 606-8502, Japan}

\email{s\underline{ }saji@u.phys.nagoya-u.ac.jp}


\maketitle

\begin{abstract}

In this paper, the Suzaku X-ray data of the Galactic Supernova Remnant (SNR) candidate G323.7$-$1.0 are analyzed to search for X-ray emission. 
Spatially-extended enhancements in the 6.4~keV line and in soft X-rays are found inside or on the radio shell.
The soft X-ray enhancement would be the hottest part of the shell-like X-ray emission along the radio shell.
The 6.4~keV line enhancement is detected at a significance level of $4.1\sigma$.
The lower limit of the equivalent width (EW) is $1.2$~keV.
The energy centroid of the 6.4~keV line is $6.40 \pm 0.04$~keV, indicating that the iron is less ionized than the Ne-like state.
If the 6.4~keV line originates from ionizing non-equilibrium thermal plasma, presence of iron-rich ejecta in a low-ionization state is required, which is disfavored by the relatively old age of the SNR.
The 6.4~keV line enhancement would be due to K-shell ionization of iron atoms in a dense interstellar medium by high-energy particles. Since there is no irradiating X-ray source, the origin of the 6.4~keV line enhancement is not likely the photoionization. The large EW can only be explained by K-shell ionization due to cosmic-ray protons with an energy of $\sim 10$~MeV, which might be generated by the shock acceleration in G323.7$-$1.0.

\end{abstract}

\section{Introduction}
Many Supernova Remnants (SNRs) show iron K$\alpha$ emission lines.
Since the energy centroids of the lines are lower than $6.7$~keV in general, the iron is less ionized than the helium-like state~\citep{2014ApJ...785L..27Y}. 
These line emissions are thought to originate from thermal processes in the ionizing plasma.
\citet{2016PASJ...68S...8S} reported a peculiar 6.4~keV line from Kes~79. 
The spatial distribution of the line emission is correlated with a molecular cloud rather than the thermal X-ray radiation. 
Thus, the 6.4~keV line emission from Kes~79 is unlikely to originate from thermal processes. 
\citet{2014PASJ...66..124S} also discovered a hint of the 6.4~keV emission line from 3C~391, which can be associated with molecular clouds.
Such a 6.4~keV line was also reported from five Galactic SNRs even though the electron temperature of their thermal plasma is lower than $\sim 1$~keV~\citep{2017_nobukawa}. Since almost no iron K-shell lines are emitted from such low-temperature plasma, the 6.4~keV line emission is thought to be generated by non-thermal processes such as interactions between low-energy cosmic-ray particles and adjacent cold gases, or photoionization.
Because of the large equivalent width (EW) of $\gtrsim 400$~eV and the absence of X-ray irradiating sources, the authors claimed that the low-energy cosmic-ray proton with the $\sim$MeV energy is the most plausible origin of the 6.4~keV line.

G323.7$-$1.0 is an SNR candidate discovered in the radio band by Molonglo Galactic Plane Survey~\citep{2014PASA...31...42G}. The radio image shows an extremely faint oval shell with a size of $\sim \timeform{51'} \times \timeform{38'}$.
Because of its large and faint radio shell, G323.7$-$1.0 would be an old SNR.
GeV and TeV gamma-ray emissions coincident with G323.7$-$1.0 have been discovered by Fermi and High Energy Stereoscopic System (H.E.S.S.) (\cite{2017ApJ...843...12A}; \cite{2015ICRC...34..886P}), respectively.
The distance to the gamma-ray source was estimated to be $\sim 5$~kpc~\citep{2017ApJ...843...12A}.
No significant X-rays have been detected from G323.7$-$1.0 so far. We, therefore, searched for X-ray emission, thermal
and/or non-thermal, from G323.7$-$1.0 with Suzaku~\citep{2007PASJ...59S...1M}. Then, we discovered spatially-extended 6.4~keV line emission. Although there is soft thermal emission which is not associated with the 6.4~keV line, 
no plasma hot enough to emit the Fe K lines is found from any region. Thus, the origin of the 6.4~keV line is due to non-thermal processes.
This paper discusses details of the 6.4~keV line emission and its origin. 
In this paper, the uncertainties are shown at the 90\% confidence level, while the errors on the data points of spectra are given at the $1\sigma$ level.

\section{Observation and data reduction
\label{section:observation_and_reduction}
}
\begin{figure}
 \begin{center}
 \FigureFile(80mm,50mm){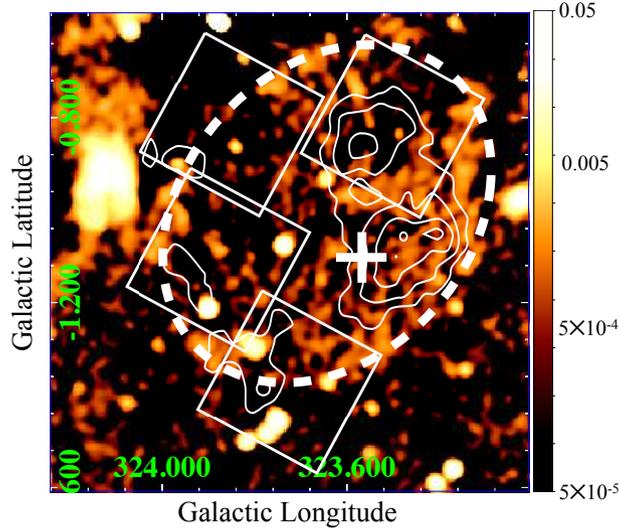}
\end{center}
  \caption{Suzaku XIS fields of view (white solid boxes) of the observations for G323.7$-$1.0 on the $843$~MHz radio map taken from SkyView (https://skyview.gsfc.nasa.gov/). The while dashed ellipse indicates the approximate shape of the radio shell. The contours show the surface brightness map of $> 1$~TeV gamma-rays~\citep{2015ICRC...34..886P}. The cross point shows the brightness peak position of the GeV gamma-ray emission~\citep{2017ApJ...843...12A}.
  \label{fig:suzaku_obs}}
\end{figure}

Suzaku observed four regions of G323.7$-$1.0. Figure~\ref{fig:suzaku_obs} shows the fields of view of the X-ray Imaging
Spectrometer (XIS) on the $843$~MHz radio image of G323.7$-$1.0.
The Suzaku observations caught the eastern part of the radio shell. 
Suzaku did not observe the brightness peaks of the GeV and TeV gamma-ray emissions as shown with the white cross point
and the contours in figure~\ref{fig:suzaku_obs}, respectively.
The details of the observations are summarized in table~\ref{table:observation_log}.

The XIS consists of four sets of CCD cameras
(XIS0-3) \citep{2007PASJ...59S..23K}.  Each CCD chip is
placed on the focal plane of the X-Ray Telescope (XRT;
\cite{2007PASJ...59S...9S}). XIS0, 2 and 3 employ
front-illuminated (FI) CCDs, while XIS1 does a
back-illuminated (BI) CCD. The CCD has 1024 by 1024 pixels, corresponding to $\timeform{17.8'} \times \timeform{17.8'}$. Since XIS2 suffered serious
damage, it has not been usable since 2006 November 9. 
In the observations of table~\ref{table:observation_log}, the XIS was operated in the normal
clocking mode with the Spaced-row Charge Injection 
\citep{2007SPIE.6686E..0QB, 2009PASJ...61S...9U}. 

\begin{table*}
\tbl{Summary of the Suzaku observations for G323.7$-$1.0.}{
\begin{tabular}{cccc}
\hline
Obs.ID & Start time (UTC) & Exposure (ks)\footnotemark[$*$] & Pointing direction ($l$, $b$)\\
\hline
 508013010 & 2013-Sep-08 10:25:42 & $36.9$ & ($\timeform{323.D8857}$, $\timeform{-1.D1033}$) \\
 508014010 & 2013-Sep-09 00:31:34 & $38.9$ & ($\timeform{323.D5092}$, $\timeform{-0.D8166}$) \\
 508015010 & 2013-Sep-09 16:05:38 & $38.8$ & ($\timeform{323.D7331}$, $\timeform{-1.D3685}$) \\
 508016010 & 2013-Sep-10 04:19:22 & $42.2$ & ($\timeform{323.D8535}$, $\timeform{-0.D8132}$) \\
\hline
\end{tabular}}
\label{table:observation_log}
\begin{tabnote}
\footnotemark[$*$] Effective exposure of the screened data.
\end{tabnote}
\end{table*}

We analyzed the archival data with the \texttt{processing version of 3.0.22.43}, using \texttt{HEAsoft version 6.21} and the calibration database (\texttt{CALDB}) released on 2016 Jun 7. 
The XIS pulse-height data for each X-ray event were converted to Pulse Invariant (PI) channel using the software \texttt{xispi} and the \texttt{CALDB}. 
The data taken at the South Atlantic Anomaly and low elevation
angles of $< \timeform{5D}$ from the night-earth rim and of $< \timeform{20D}$
from the day-earth rim were excluded.
The ancillary response files and the redistribution matrix files were produced using  \texttt{xissimarfgen} and \texttt{xisrmfgen} packages of \texttt{HEAsoft}, respectively.

\section{Analysis and results}
 \label{section:analysis}

\subsection{Image}
\label{subsection:image}

In order to search for X-ray sources in the G323.7$-$1.0 field,
XIS images in the soft ($0.5$--$3$~keV), hard ($5$--$8$~keV) and $6.3$--$6.5$~keV bands were made. 
XIS1 (BI-CCD) data were not used to make the images above $5$~keV, since the signal-to-noise ratio in the hard band is worse than that of the FI-CCDs.
The corners of the CCD chips illuminated by the $^{55}$Fe calibration sources were excluded from the images above $5$~keV.
The Non-X-ray Backgrounds (NXBs) were estimated from the data within $\pm 300$~days of the observation using \texttt{xisnxbgen}~\citep{2008PASJ...60S..11T},
and were subtracted from the images. 
After the NXB subtraction, each image was divided by an exposure map simulated using the XRT$+$XIS simulator \texttt{xissim}~\citep{2007PASJ...59S.113I} for vignetting corrections. 

\begin{figure}
 \begin{center}
 \FigureFile(160mm,50mm){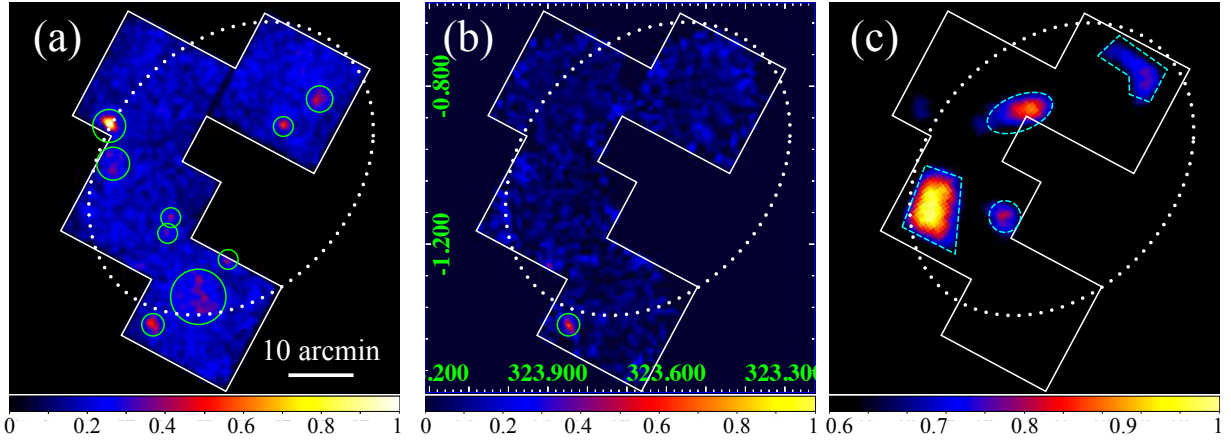}
 \end{center}
  \caption{XIS images of G323.7$-$1.0 in the (a) $0.5$--$3$~keV, (b) $5$--$8$~keV and (c) $6.3$--$6.5$~keV bands. The NXBs are subtracted and then the images are corrected the vignetting effect. The images (a) and (b) are binned with $8 \times 8$~pixels and then smoothed with a Gaussian function of $\sigma = 5$~bins. The image (c) is binned with $32 \times 32$~pixels and then smoothed with a Gaussian function of $\sigma = 10$~bins. The axes of the Galactic longitude and latitude are plotted on image (b). The largest dotted ellipse in each panel indicates the approximate shape of the radio shell (see figure~\ref{fig:suzaku_obs}).}
  \label{fig:image}
\end{figure}

Three panels in figure~\ref{fig:image} show the images thus obtained in the soft, hard and $6.3$--$6.5$~keV bands.
The soft and hard-band images are binned with $8 \times 8$ pixels and then smoothed with a Gaussian function of $\sigma = 5$~bins. The $6.3$--$6.5$~keV band image is binned with $32 \times 32$ pixels and smoothed with a Gaussian function of $\sigma = 10$~bins.

In the soft-band image, there are several point-like sources marked with solid small circles, while
a diffuse structure, indicated by the largest solid circle with a diameter of $\timeform{8.4'}$, is found in the southern field. 
This diffuse structure is called ``the soft source" hereafter.
On the other hand, there is no prominent structure in the hard band except for one point source that is also seen in the soft band.
In the $6.3$--$6.5$~keV band, some structures are seen as marked with dashed curves.
Because of their angular sizes of $\gtrsim 5'$, which is larger than the half-power diameter of XRT ($\sim 2'$), they should not be point sources, but diffuse sources.
These sources are hereafter called ``the 6.4~keV clumps''.

\subsection{Spectrum}
\label{subsection:Spectrum}

\subsubsection{The 6.4~keV clumps}
In order to investigate the 6.4~keV clumps, the ``enhance region" was defined as illustrated with dashed curves in figure~\ref{fig:image}c. 
The whole field excluding the soft source, the point sources and the enhance region was defined as the reference region.
XIS0 and 3 spectra were extracted from the enhance and reference regions. 
 Although the enhance region consists of four parts, the data extracted from all areas were added to improve the statistics.
The NXBs were estimated using the data within $\pm300$~days of the observation, and were subtracted from the enhance and reference spectra.

Since G323.7$-$1.0 is located on the Galactic plane ($b = \timeform{-1.D0}$), there should be emission lines of the Galactic Ridge X-ray Emission (GRXE). The GRXE exhibits lines of neutral iron (Fe I) K$\alpha$, helium-like iron (Fe XXV) K$\alpha$, hydrogen-like iron (Fe XXVI) K$\alpha$ and Fe I K$\beta$ at $6.40$~keV, $6.68$~keV, $6.97$~keV and $7.06$~keV, respectively~(e.g.,~\cite{2016PASJ...68...59Y}). Thus we fitted the $5$--$8$~keV band spectra with a model consisting of a power-law and four Gaussian functions. The energy centroids of the Gaussian functions were fixed to be $6.40$~keV, $6.68$~keV, $6.97$~keV and $7.06$~keV. The intensity of the Fe I K$\beta$ line was set to $0.125$ times that of the Fe I K$\alpha$ line~\citep{1993A&AS...97..443K}, while those of the other lines were free parameters. The width of the Gaussian functions were fixed to be zero, except for the Fe XXV K$\alpha$ line. Since the Fe XXV K$\alpha$ line is a blend of the resonance, inter-combination and forbidden lines, the line width was fixed to be $23$~eV based on~\citet{2007PASJ...59S.245K}. 
Both of the photon index and the normalization of the power-law function were allowed to vary.

\begin{figure*}
 \begin{center}
 \FigureFile(80mm,80mm){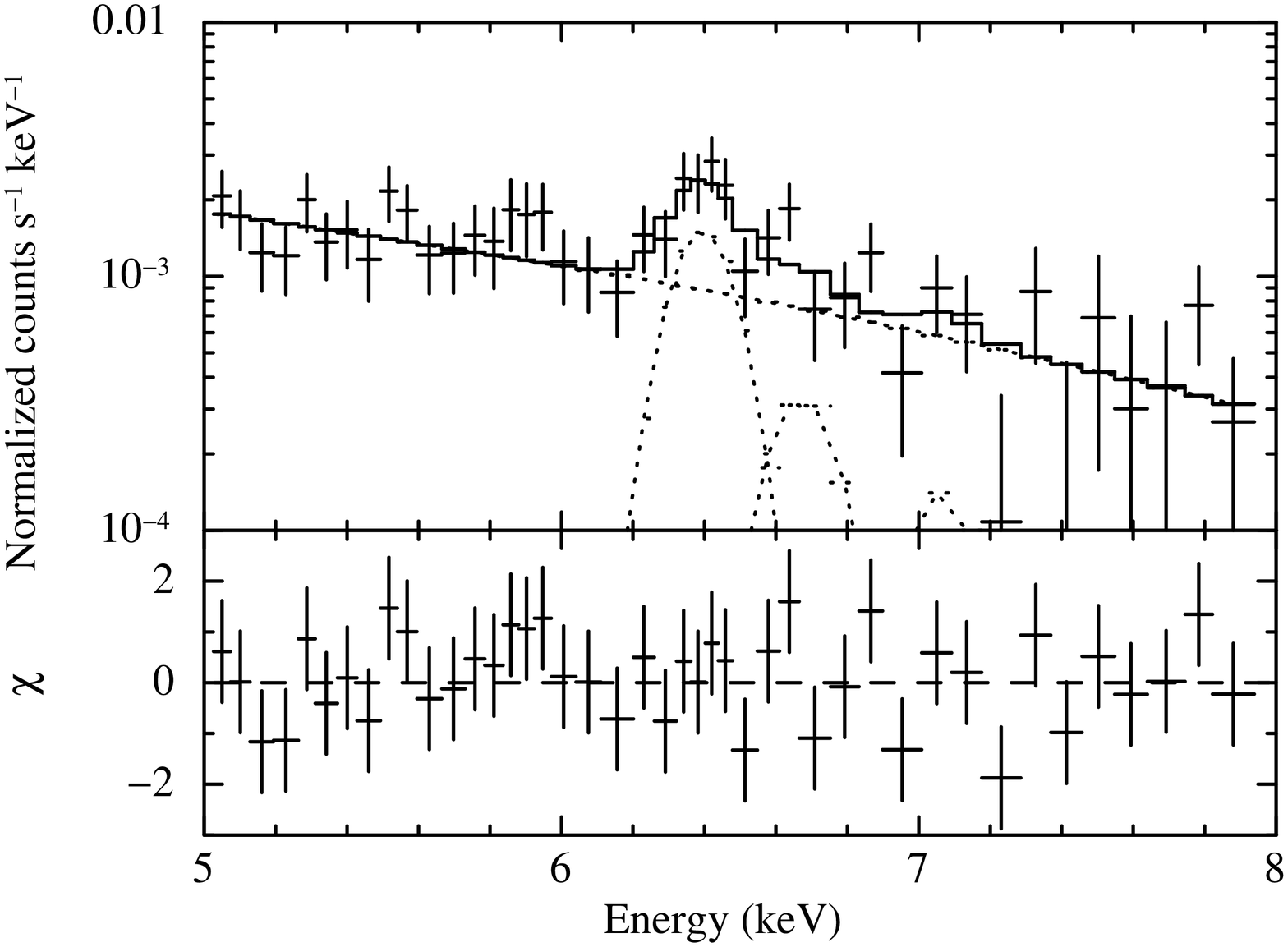}
 \FigureFile(80mm,80mm){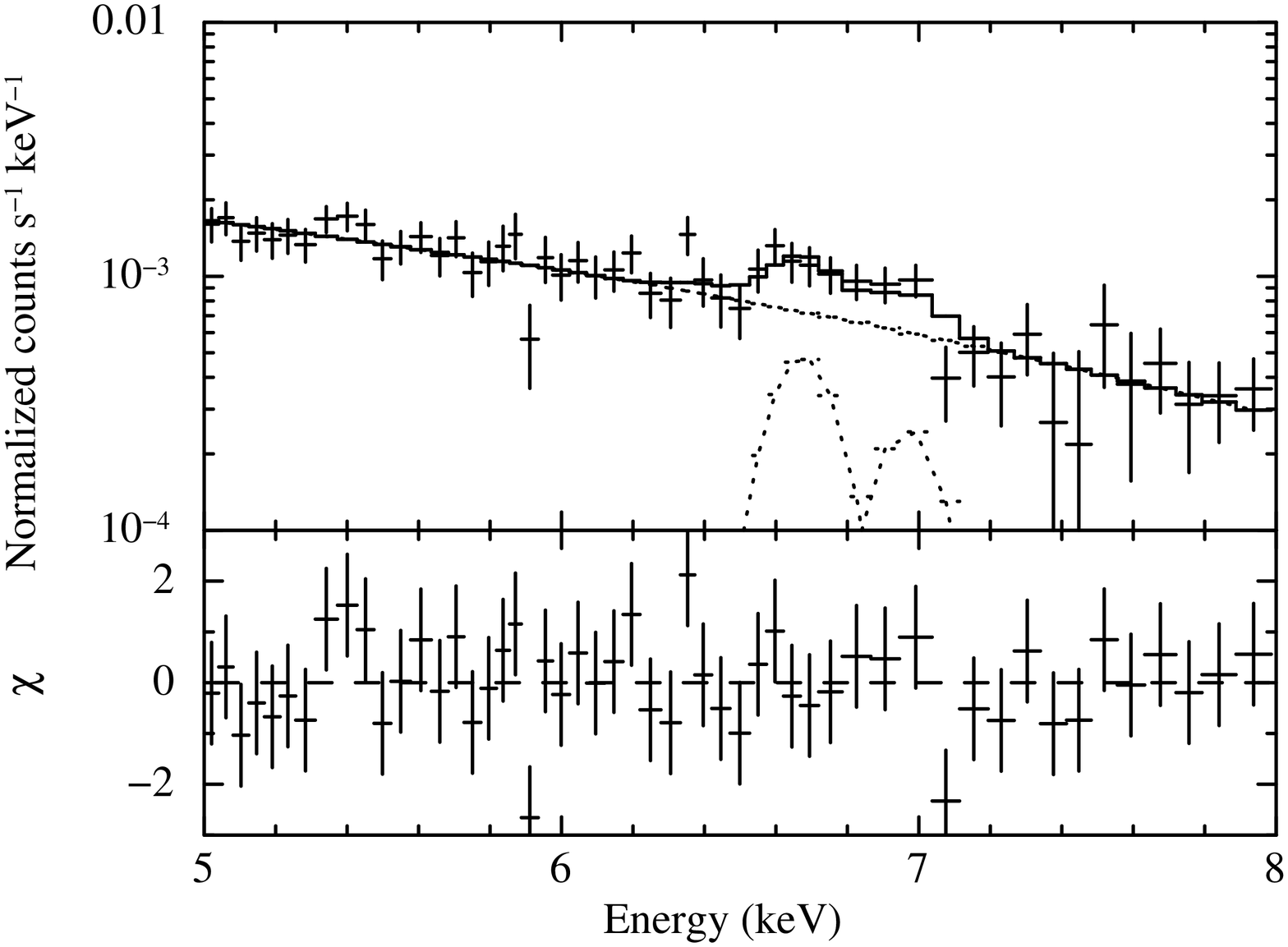} 
 \end{center}
  \caption{Top panels: XIS0 $+$ 3 spectra extracted from the enhance region (left) and the reference region (right) fitted with the the model consisting a power-law and four Gaussian functions model. The best-fit models are plotted with the solid lines, while each component is plotted with the dotted line. The vertical axes are normalized by the ratio of the area including the vignetting correction. Bottom panels:  Residuals between the data and the best-fit models.}
  \label{fig:spec_fit}
\end{figure*}

Figures~\ref{fig:spec_fit} show the spectra extracted from the enhance region and the reference region with the best-fit models.
Both of the model fittings are statistically acceptable with $\chi^2_{\rm red} = 0.84$ ($38$ d.o.f.) for the enhance region and $\chi^2_{\rm red} = 0.88$ ($46$ d.o.f.) for the reference region.
The best-fit parameters are summarized in table~\ref{table:fit_param}.
The 6.4~keV line intensity of the enhance region is significantly higher than that of the reference region. The significance level of the excess is $4.1\sigma$.
On the other hand, the other parameters are consistent between the two regions.

\begin{figure}
 \begin{center}
 \FigureFile(80mm,50mm){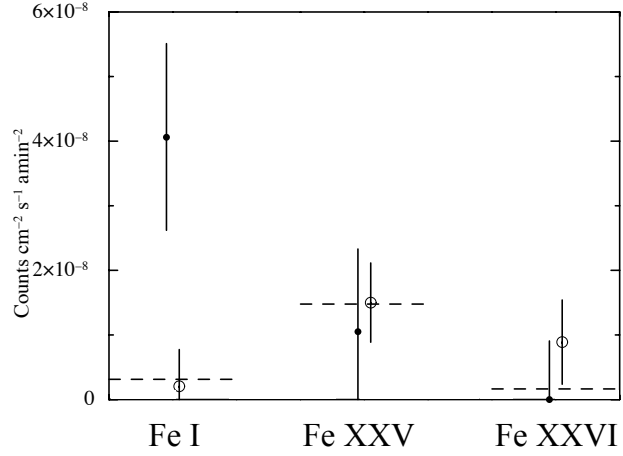}
 \end{center}
  \caption{Intensities of the Fe I (left), Fe XXV (center) and Fe XXVI (right) K$\alpha$ lines of the enhance region and the reference region, compared with that of the GRXE. The filled and open circles indicate the line intensity estimated from the spectral fitting for the enhance region and the reference region, respectively. The dashed horizontal lines indicate the line intensities of the GRXE calculated based on \citet{2013PASJ...65...19U} and \citet{2016PASJ...68...59Y}.}
  \label{fig:line_int}
\end{figure}

Next we estimated the line intensities of the GRXE at the position of G323.7$-$1.0, $(l, b) = (\timeform{323.D7}, \timeform{-1.D0})$.
Using the longitude profile of \citet{2013PASJ...65...19U} and the latitude profile of \citet{2016PASJ...68...59Y}, the GRXE line intensities are calculated to be $3.1 \times 10^{-9}$ (Fe I K$\alpha$), $1.5 \times 10^{-8}$ (Fe XXV K$\alpha$) and $1.6 \times 10^{-9}$ (Fe XXVI K$\alpha$)~photons~cm$^{-2}$~s$^{-1}$~arcmin$^{-2}$.
Figure~\ref{fig:line_int} shows the comparison of the line intensities between the enhance region, the reference region and the GRXE model. The intensity of the 6.4~keV line of the enhance region is significantly higher than the Fe I K$\alpha$ line of the GRXE, while that of the reference region is consistent with the GRXE. The other two lines are almost consistent with the GRXE model, although the Fe XXVI K$\alpha$ intensity of the reference region is slightly higher than the model.
Thus the 6.4~keV line enhancement cannot be explained by the GRXE.

\begin{table*}[bhtp]
\tbl{Best-fit parameters of the model fitting for the spectra of the enhance and reference regions.}{
\begin{tabular}{clcc}
\hline
\multicolumn{2}{c}{Parameter} &  \multicolumn{2}{c}{Best-fit values} \\
 & & Enhance region & Reference region \\
\hline
Power-law & Photon index & $2.2^{+1.0}_{-0.9}$ & $2.0 \pm 0.5$ \\
 & Intensity\footnotemark[$*$] & $(3.3 \pm 0.5) \times 10^{-15}$ & $(3.2 \pm 0.3) \times 10^{-15}$ \\
 \hline
Line intensities\footnotemark[$\dagger$] & Fe I K$\alpha$ ($6.40$~keV) & $4.1 \pm 1.4$ & $< 0.8$ \\
 & Fe XXV K$\alpha$ ($6.68$~keV) & $< 2.3$ & $1.5 \pm 0.6$ \\
 & Fe XXVI K$\alpha$ ($6.97$~keV) & $< 0.9$ & $0.9 \pm 0.7$ \\
 \hline
\end{tabular}}
\label{table:fit_param}
\begin{tabnote}
 \footnotemark[$*$] Observed intensity in the $5$--$8$~keV band in units of erg cm$^{-2}$ s$^{-1}$ arcmin$^{-2}$.\\
 \footnotemark[$\dagger$] Photon intensities in units of $10^{-8}$ photons cm$^{-2}$ s$^{-1}$ arcmin$^{-2}$.
\end{tabnote}
\end{table*}

\begin{figure}
 \begin{center}
 \FigureFile(80mm,50mm){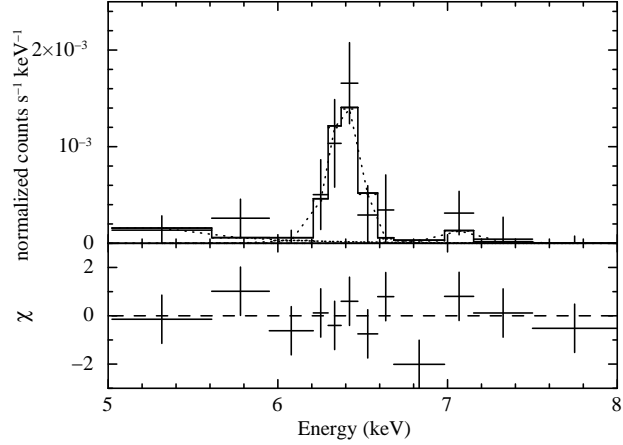}
 \end{center}
  \caption{Top panel: $5$--$8$~keV spectrum of the enhance region fitted with the model consisting of a power-law and two Gaussian functions model. The spectrum of the reference region is subtracted after the area correction. The best-fit model is shown with the solid line, while each component is plotted with the dotted line. Bottom paned: Residuals between the data and the best-fit model.}
  \label{fig:spec_eqwidth}
\end{figure}

Then, we calculated the EW of the 6.4~keV line enhancement. 
To estimate the enhanced component, the spectrum of the reference region was subtracted from that of the enhance region after correcting the difference of the area taking account of the vignetting.
Figure~\ref{fig:spec_eqwidth} shows the spectrum of the enhance region after the subtraction. The spectrum was fitted with a model consisting of a power-law and two Gaussian functions. The energy centroids of the Gaussian functions were fixed to be $6.40$~keV and $7.06$~keV assuming the Fe I K$\alpha$ and the Fe I K$\beta$ lines, respectively. The intensity of the Fe I K$\beta$ line was set to $0.125$ times that of the Fe I K$\alpha$ line. The width of each Gaussian function was fixed to be zero.  Both of the photon index and the normalization of the power-law component were free parameters.
The flux of the power-law component is consistent with zero.
The lower limit of the EW of the 6.4~keV line is $2.3$~keV with a confidence level of $90$~\%. 
The continuum level and hence the EW are sensitive to the vignetting correction.
The lower limit of the EW of the 6.4~keV line is $1.2$~keV when the uncertainty of the vignetting correction is assumed to be $5$\% according to {\it Suzaku technical description}\footnote[1]{https://heasarc.gsfc.nasa.gov/docs/suzaku/prop\_tools/suzaku\_td}.
When the energy centroid of the Gaussian function for the 6.4~keV line is allowed to vary, it becomes $6.40 \pm 0.04$~keV. The energy centroid is consistent with that of the Fe I K$\alpha$ line.

\subsubsection{The soft source} \label{sec:soft}
\begin{figure}
 \begin{center}
 \FigureFile(80mm,80mm){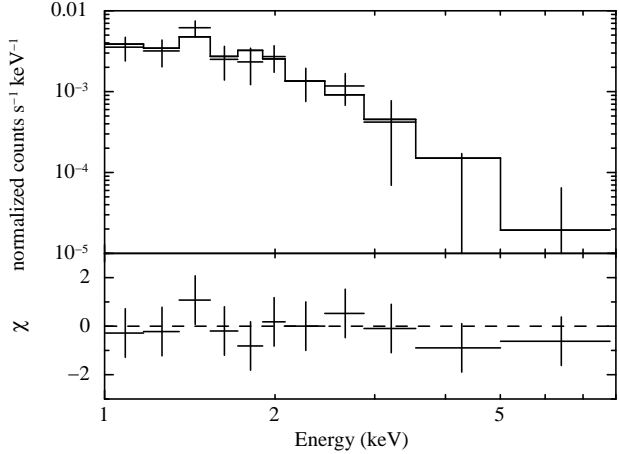}
  \caption{Top panel: Background subtracted spectrum of the soft source in the $1$--$8$~keV band fitted with the absorbed thin-thermal plasma model. The solid line shows the best-fit model. Bottom panel: Residuals between the data and the best-fit model.
  }
  \label{fig:spec_soft}
 \end{center}
\end{figure}

In order to investigate the soft source, we extracted an XIS0 $+$ 3 spectrum from the largest solid circle in figure~\ref{fig:image}a. 
The spectrum of the reference region was used as the background.
Figure~\ref{fig:spec_soft} shows the background-subtracted spectrum. 

Although the spectrum has poor statistics, a hint of an emission line can be seen at $\sim 1.4$~keV. We, then, tried fitting the spectrum with an optically-thin thermal plasma model ({\tt APEC}) suffering from an interstellar absorption ({\tt phabs}) fixing the metal abundance of a solar value~\citep{2003ApJ...591.1220L}. 
The model represents the spectrum well with the $\chi^2_{\rm red} = 0.44$ ($8$ d.o.f.).
We obtained the plasma temperature $kT=1.1^{+0.5}_{-0.3}$~keV and the absorption column density $N_{\rm H}=(1.8\pm0.6)\times 10^{22}$~cm$^{-2}$ as well as the normalization $2.8^{+1.8}_{-1.1}\times 10^{-4}$.
The normalization is the volume emission measure represented by ${10^{-14} n_{\rm e}n_{\rm H} V/{4\pi d^2}}$, where $n_{\rm e}$, $n_{\rm H}$, $V$, and $d$ are the electron density, the hydrogen density, the plasma volume, and the distance from the solar system, respectively.
The observed $1$--$8$~keV flux is estimated to be $8.4\times 10^{-14}$~erg~s$^{-1}$~cm$^{-2}$.

\section{Discussion} \label{section:discussion}
\subsection{The soft source}
We constrained the interstellar absorption toward the soft source of $N_{\rm H}=(1.8 \pm 0.6)\times 10^{22}$~cm$^{-2}$. 
Assuming an averaged number density of $1$H~cm$^{-3}$, the absorption column density corresponds to a distance of $6 \pm 2$~kpc,
which is consistent with the distance $d=5$~kpc of the gamma-ray source coincident with G323.7$-$1.0 \citep{2017ApJ...843...12A}.
Therefore, the soft source possibly belongs to G323.7$-$1.0, and is a thermal plasma generated by the shock heating.
The source radius is calculated to be 6\,($d$/5~kpc)~pc based on the angular radius of $\sim 4.\!'2$.
Using the plasma temperature and the volume emission measure (see section~\ref{sec:soft}), 
the density and the total mass are calculated to be $0.06$~cm$^{-3}$ and $1.9$~$M_{\odot}$, respectively.
Here, the plasma volume is estimated assuming a sphere with the radius of $6$~pc.

The soft source is located near the southeastern shell of G323.7$-$1.0 (\cite{2014PASA...31...42G}, figure~\ref{fig:image}a).
The size of the soft source is $\sim \timeform{8.'4}$, which is far smaller than the entire size of the radio shell of G323.7$-$1.0 ($\sim \timeform{51'} \times \timeform{38'}$;~\cite{2014PASA...31...42G}).
If G323.7$-$1.0 is an old SNR and the age $t$ is larger than $10000$~yr, the plasma temperature is roughly estimated to be 0.3 ($t$/$10^4$~yr)$^{-6/5}$~keV using the Sedov solution assuming the explosion energy of $1 \times 10^{51}$~erg and the number density of the interstellar matter of $1$H~cm$^{-3}$. Soft X-rays from such low temperature plasma are difficult to be detected because of the interstellar absorption ($N_{\rm H}\sim 2\times 10^{22}$~cm$^{-2}$). 
Shock heated plasma would make a shell-like soft X-ray emission along the radio shell. The soft source would be the hottest part of the shell emission, and the other regions would be unseen due to the large interstellar absorption.

\subsection{The 6.4~keV clumps}
Thermal plasma in a non-equilibrium ionizing (NEI) state can emit the iron K-shell line at $\sim 6.4$~keV. 
When we assume the solar abundance of~\citet{2003ApJ...591.1220L} and the electron temperature of $< 5$~keV, the EW of the iron K$\alpha$ line from the NEI plasma is calculated to be smaller than $\sim 300$~eV using the NEI code in {\tt XSPEC}~\citep{1996ASPC..101...17A}.
The iron abundance higher than $\sim 4$~solar is required to explain the observed EW. Such high abundance can only be realized by iron-rich ejecta.
The upper limit of the observed line centroid is 6.44~keV, corresponding to an ionization degree of the Ne-like state~(\cite{1993ApJ...409..846B}; \cite{2003A&A...410..359P}; \cite{2004A&A...414..377M}).
If the 6.4~keV line comes from the NEI plasma, the ionization timescale is estimated to be $t \lesssim 300 (n_{\rm e}/1{\rm ~cm}^{-3})^{-1}$~yr using the NEI code. Here, $n_{\rm e}$ is the electron number density in the NEI plasma.
The ejecta in G323.7$-$1.0 cannot be in such a low-ionization state since G323.7$-$1.0 has a large and extremely faint radio shell and hence would not be a young SNR. Indeed, \citet{2014ApJ...785L..27Y} reported that only young SNRs exhibit such a low-ionization state using the Suzaku archive data of many SNRs. Also, the oldest SNR which shows the energy centroid of $< 6.44$~keV is RCW~86, the remnant of SN~185~\citep{2011ApJ...741...96W}. 
Thus, the thermal plasma is unlikely as the origin of the observed 6.4~keV~line.

Then, the 6.4~keV line emission is explained by non-thermal processes such as K-shell ionization by photons, non-thermal electrons or protons with the energy above the Fe K-edge (7.1~keV). In the following, we discuss which particle is the dominant source for the observed 6.4~keV line.

In the case of the photoionization, a bright X-ray source is required in the vicinity of G323.7$-$1.0.
Cir~X-1, which is an X-ray binary hosting a neutron star, is located $\sim 2$~degrees away from G323.7$-$1.0.
The distance from the solar system is $4.1$~kpc \citep{2005ApJ...619..503I}.
Thus, the minimum distance between G323.7$-$1.0 and Cir~X-1 is $4.1\,{\rm sin} (\timeform{2D})$~kpc $= 140$~pc.
Since the angular size of the largest 6.4~keV clump is $\sim 8'$, the actual size of the clump is $\sim 10$~pc assuming that the distance from the solar system to G323.7$-$1.0 is the same as that to Cir~X-1, $4.1$~kpc.
Then, the hydrogen number density of the clump is estimated to be $\sim 600$~cm$^{-3}$ based on an empirical relation between the density and size of $n_{\rm H} = 180(D/40 {\rm pc})^{-0.9}$~cm$^{-3}$~\citep{1987ApJS...63..821S}, where $D$ is the size of the clump. Note that there are no available data of $^{12}$CO or HI with enough angular resolution to estimate the density of the clump.
Using the photon index of Cir~X-1 ($\Gamma = 2.6$;~\cite{2005ApJ...619..503I}) and the cross-sections for the photoionization~\citep{1992ApJ...400..699B}, a luminosity of $\sim 2.2 \times 10^{39}$~erg s$^{-1}$ in the $0.1$--$100$~keV band is required for Cir~X-1 to generate the observed 6.4~keV line intensity. Since the luminosity exceeds the Eddington luminosity for a neutron star ($\sim 2 \times 10^{38}$~erg~s$^{-1}$) by an order of magnitude, the photoionization is unlikely to be the significant origin of the 6.4~keV line. 

A scenario that the iron in the cold and dense cloud is ionized by electrons with an energy of $10$--$100$~keV might also be possible (e.g., \cite{2011PASJ...63..535D}; \cite{2015ApJ...807L..10N}).
In this process, strong continuum emission must be generated by bremsstrahlung.
The EW of the iron K$\alpha$ line for the bremsstrahlung is $\sim 400$~eV at most assuming a solar abundance~\citep{2011PASJ...63..535D}. Since the lower limit of the EW of the observed 6.4~keV line is $1.2$~keV, an anomalous large iron abundance of 3~solar or more is required. 

The 6.4~keV line can also be produced via ionization by protons with an energy of $\sim 10$~MeV. 
In this process, the EW of the Fe I K$\alpha$ emission is $\sim 1$~keV or higher for gas with a solar abundance~\citep{2011PASJ...63..535D}. 
Thus the observed EW ($>1.2$~keV) can be well explained with the proton origin.
The Fe I K$\alpha$ intensity is calculated by
\begin{equation}
\label{calc_line_intensity}
I_{6.4{\rm keV}} = \frac{1}{4\pi}\int \sigma_{6.4{\rm keV}}\, v\, n_{\rm H}\, l\, Z_{\rm Fe}\, \frac{dN}{dE}\, dE,
\end{equation}
where $\sigma_{6.4{\rm keV}}$ is the proton's cross section for the Fe I K$\alpha$ emission~\citep{2012A&A...546A..88T}, $v$ is the velocity of the proton, $n_{\rm H}$ is the hydrogen number density of the clump, $l$ is the thickness of the clump along the line-of-sight, $Z_{\rm Fe}$ is the iron abundance of the clump, and $dN/dE$ is the spectral number distribution of the proton.
Since the angular size of the largest 6.4~keV clump is $\sim 8'$, the thickness of the clump along the line-of-sight is assumed to be $l = d\, {\rm sin}(8')$. Here $d$ is the distance from the solar system to the clump. 
When the spectrum of the proton is set to be $dN/dE \propto E^{-2.7}$ assuming the Galactic cosmic-rays~\citep{1999RvMPS..71..165C}, a proton energy density of $\sim 300 (100~{\rm cm}^{-3}/n_{\rm H})(5~{\rm kpc}/d)$~eV~cm$^{-3}$ in the $0.1$--$1000$~MeV band is required to generate the observed 6.4~keV line intensity. 
Here the abundance of the iron $Z_{\rm Fe}$ is assumed to be that of the Sun~\citep{2003ApJ...591.1220L}.
This energy density is much higher than the canonical value for the Galactic cosmic-ray ($\sim 1$~eV cm$^{-3}$; \cite{2012PhRvL.108e1105N}). Those protons are thought to be accelerated in G323.7$-$1.0.

The 6.4 keV clumps are located around or inside the faint radio shell (figure~\ref{fig:image}c).
The shell would be the diffusive shock acceleration cite of protons.
Higher-energy protons easily escape from the acceleration site (the SNR shell region), 
since their mean free paths are longer than those of lower-energy ones~\citep{2011MNRAS.410.1577O}.
Then the escaped high-energy protons would hit a high-density cloud located far from the shell and produces high-energy gamma-rays.
In fact, the brightest peak of the gamma-ray source is about $20'$ away from the 6.4~keV 
clumps (see figures~\ref{fig:suzaku_obs} and \ref{fig:image}c).
The intensities of the $6.4$~keV line and the high-energy gamma-rays depend on the density of low-energy protons and high-energy protons, respectively, as well as neutral gas density. More qualitative discussions on possible correlations between the 6.4~keV line and the gamma-ray intensities require detailed studies on the 6.4~keV line, gamma-rays and $^{12}$CO or HI with an angular resolution better than a few arcmin. These would be future projects.

\section*{Acknowledgments}
SS is supported by Grant-in-Aid for Japan Society for the Promotion of Science (JSPS) Fellows number JP17J11471.
This work is also supported by MEXT KAKENHI Grant Number JP15H02070 (HM), JP25887028 (HU), JP15H02090, JP17K14289 (MN), and JSPS Number JP16J00548 (KKN).
KKN is supported by Research Fellowships of JSPS for Young Scientists.

\end{document}